\documentclass[10pt,a4]{article}
\title{Two Dimensional Angle of Arrival Estimation}
\author{Santhosh Kumar, Pradip Sircar\\
	\small Department of Electrical Engineering\\
	\small Indian Institute of Technology Kanpur\\
	\small Kanpur 208016, Uttar Pradesh, India}

\usepackage{graphicx}
\usepackage{amssymb}	
\usepackage{amsmath}
\usepackage{pmat}

\date{}
\begin{document}

\maketitle

\section{Introduction}
\label{section:introduction}

We present a new method for the estimation of two dimensional (2D) angles of arrival (AOAs), namely, azimuth and incidence angles of multiple narrowband signals of same frequency in the far field of antenna array. In this report, we propose an algorithm which uses an L-shape array configuration ~\cite{Tayem}. We extend the polynomial based approach presented in
~\cite{Singh} to two dimensional AOA estimation.

We Consider an L-shaped antenna array as shown in Figure 1 in the XZ plane ~\cite{Tayem}.

\begin{figure}[h!]
  \centering
    \includegraphics[width=2.5in]{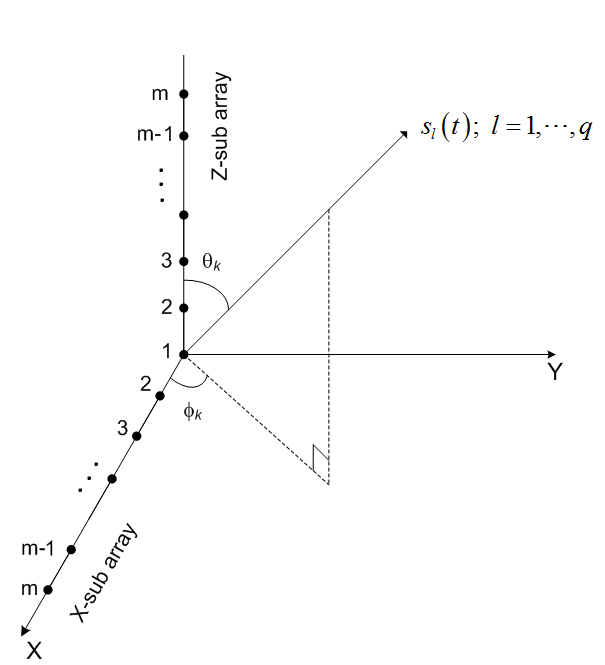}
    \caption{The L-shaped array configuration}
\label{fig:array}
\end{figure}

The gain of an element at a position $(x,y,z)$ for a signal with reference to origin from the direction $(\theta, \phi)$ is given by
\begin{equation}
a(\theta, \phi) = e^{j(2\pi/\lambda)(x \sin{\theta}\cos{\phi} + y \sin{\theta}\sin{\phi} + z \cos{\theta})}
\end{equation}

Suppose that there are $q$ narrow band sources, with same wavelength $\lambda$ impinging on each sensor. An $l$th source has an incidence angle $\theta_{l}$ and an azimuth angle $\phi_{l}$, $l=1,2,...,q$.

The signal received at the $i$th element of the Z-subarray is given by ~\cite{Tayem}
\begin{equation}
z_i(t)=\sum_{l=1}^{q}a_{z,i}(\theta_{l},\phi_{l})s_{l}(t)+n_{z,i}(t)
\end{equation}
where
\begin{eqnarray}
a_{z,i}(\theta_{l},\phi_{l})&=&e^{j\beta d\cos{ \theta_{l}} } \nonumber\\
&=&e^{j2\pi (i-1)(d/\lambda)\cos{\theta_{l}}}\nonumber\\ &=&e^{j(i-1)\psi_{l}} \nonumber
\end{eqnarray}
\begin{equation}
\psi_{l}=2\pi(d/\lambda)\cos{\theta_{l}}
\label{define_psi}
\end{equation}
$s_{l}(t)$ \space is the complex envelope of the $l$th signal arriving from the direction $(\theta_{l},\phi_{l})$\\
and $n_{z,i}(t)$ is the zero-mean additive white Gaussian noise (AWGN) whose properties are mentioned in Section 2.2.

The signal received at the $i$th element of the X-subarray is given by
\begin{equation}
x_i(t)=\sum_{l=1}^{q}a_{x,i}(\theta_{l},\phi_{l})s_{l}(t)+n_{x,i}(t)
\end{equation}
where
\begin{eqnarray}
a_{x,i}(\theta_{l},\phi_{l})&=&e^{j2\pi (i-1)(d/\lambda)\sin{\theta_{l}}\cos{ \phi_{l}} }\nonumber\\
&=&e^{j(i-1)\xi_{l}},\nonumber
\end{eqnarray}
\begin{equation}
\xi_{l}=2\pi(d/\lambda)\sin{\theta_{l}}\cos{\phi_{l}}
\label{define_varphi}
\end{equation}
and $n_{x,i}(t)$ is the zero-mean AWGN.

If we collect outputs of all the sensors on the Z-subarray  into an $m\times1$ vector, we have
\begin{equation}
\mathbf{z}(t)=\sum_{l=1}^{q}\mathbf{a}_{z}(\theta_{l},\phi_{l})s_{l}(t)+\mathbf{n}_{z}(t)
\label{z_vector_in_sum}
\end{equation}
where
\begin{equation}
\mathbf{z}(t)= [ z_{1}(t) \hspace{.2in}z_{2}(t)\hspace{.1in} \cdots\hspace{.1in}  z_{m}(t) ]^{T}\nonumber
\end{equation}
$\mathbf{a}_{z}(\theta_{l},\phi_{l})$ \hspace{.05in} is the $m\times1$ steering vector of the Z-subarray towards the direction $(\theta_{l},\phi_{l})$ given by
\begin{equation}
\mathbf{a}_{z}(\theta_{l},\phi_{l})=[a_{z,1}(\theta_{l},\phi_{l}\hspace{0.2in}a_{z,2}(\theta_{l},\phi_{l})\hspace{0.1in}\cdots\hspace{0.1in}a_{z,m}(\theta_{l},\phi_{l})]^T\nonumber
\end{equation}
and
\begin{equation}
\mathbf{n}_{z}(t)=
\left[\begin{array}{cccc} 
n_{z,1}(t) 	& n_{z,2}(t)	 &\cdots	&n_{z,m}(t)
\end{array}\right]^T\nonumber
\end{equation}

We can rewrite (\ref{z_vector_in_sum}) as
\begin{equation}
\small{
\mathbf{z}(t)=\left(\begin{array}{cccc}
 a_{z,1}(\theta_{1},\phi_{1})& a_{z,1}(\theta_{2},\phi_{2})  & \cdots & a_{z,1}(\theta_{q},\phi_{q}) \\ 
 a_{z,2}(\theta_{1},\phi_{1})& a_{z,2}(\theta_{2},\phi_{2})&  \cdots  &a_{z,2}(\theta_{q},\phi_{q})  \\ 
 \vdots&\vdots  &\vdots  &\vdots  \\ 
 a_{z,m}(\theta_{1},\phi_{1})&a_{z,m}(\theta_{2},\phi_{2})  &\cdots  &a_{z,m}(\theta_{q},\phi_{q}) 
\end{array}\right)   
\left(\begin{array}{c}
s_{1}(t)\\
s_{2}(t)\\
\vdots \\
s_{q}(t) \end{array}\right)+
\left(\begin{array}{c}
n_{z,1}(t)\\
n_{z,2}(t)\\
\vdots\\
n_{z,m}(t)
 \end{array}\right)}
\label{z_vector_in_matrix}
\end{equation}

Similarly for the X-subarray, we have
\begin{equation}
\mathbf{x}(t)=\sum_{l=1}^{q}\mathbf{a}_{x}(\theta_{l},\phi_{l})s_{l}(t)+\mathbf{n}_{x}(t)
\label{x_vector_in_sum}
\end{equation}
where
\begin{equation}
\mathbf{x}(t)= [ x_{1}(t) \hspace{.2in}x_{2}(t)\hspace{.1in} \cdots\hspace{.1in}  x_{m}(t) ]^{T}\nonumber
\end{equation}

\begin{equation}
\mathbf{a}_{x}(\theta_{l},\phi_{l})=[a_{x,1}(\theta_{l},\phi_{l}\hspace{0.2in}a_{x,2}(\theta_{l},\phi_{l})\hspace{0.1in}\cdots\hspace{0.1in}a_{x,m}(\theta_{l},\phi_{l})]^T\nonumber
\end{equation}
and
\begin{equation}
\mathbf{n}_{x}(t)=[n_{x,1}(t)\hspace{.2in}n_{x,2}(t)\hspace{.1in}\cdots\hspace{.1in}n_{x,m}(t)]^T\nonumber
\end{equation}

Note that (\ref{x_vector_in_sum}) can be rewritten as
\begin{equation}
\small{
\mathbf{x}(t)=\left(\begin{array}{cccc}
 a_{x,1}(\theta_{1},\phi_{1})& a_{x,1}(\theta_{2},\phi_{2})  & \cdots & a_{x,1}(\theta_{q},\phi_{q}) \\ 
 a_{x,2}(\theta_{1},\phi_{1})& a_{x,2}(\theta_{2},\phi_{2})  & \cdots & a_{x,2}(\theta_{q},\phi_{q}) \\ 
 \vdots			     & \vdots  			     &\vdots  & \vdots 			     \\ 
 a_{x,m}(\theta_{1},\phi_{1})& a_{x,m}(\theta_{2},\phi_{2})  &\cdots  & a_{x,m}(\theta_{q},\phi_{q}) 
\end{array}\right)   
\left(\begin{array}{c}
s_{1}(t)\\
s_{2}(t)\\
\vdots \\
s_{q}(t)
\end{array}\right)+
\left(\begin{array}{c}
n_{x,1}(t)\\
n_{x,2}(t)\\
\vdots\\
n_{x,m}(t)
\end{array}\right)}
\label{x_vector_in_matrix}
\end{equation}

\section{Proposed Method}
Consider the Z-subarray:

On substituting the values of $a_{z,i}(\theta_{l},\phi_{l})$ in     (\ref{z_vector_in_matrix}) for $i=1,2,...,m$ and $l=1,2,..,q$ , we have
\begin{equation}
\mathbf{z}(t)=
\left(\begin{array}{cccc}
1 &   1 &  \cdots &    1 \\
e^{j\psi_{1}} & e^{j\psi_{2}} & \cdots & e^{j\psi{q}} \\
\vdots &  \vdots &   \vdots & \vdots  \\
e^{j(m-1)\psi_{1}} &e^{j(m-1)\psi_{2}}&\cdots &e^{j(m-1)\psi_{q}}
\end{array}\right)
\left(\begin{array}{c}
s_{1}(t)\\
s_{2}(t)\\
\vdots \\
s_{q}(t)
\end{array}\right) +
\left(\begin{array}{c}
n_{z,1}(t)\\
n_{z,2}(t)\\
\vdots    \\
n_{z,m}(t)
\end{array}\right)
\end{equation}
where $\lbrace\psi_{l}\rbrace$ are defined in (\ref{define_psi}).

By collecting $M$ equispaced snapshots from each sensor, a data matrix Z is formed as follows
\begin{equation} 
\label{eq:Z_Data_Matrix}
\begin{split}
Z=
\left(\begin{array}{cccc}
1			& 1			& \cdots 	& 1		\\
e^{j\psi_{1}}   	& e^{j\psi_{2}} 	& \cdots        & e^{j\psi_{q}} \\
\vdots 			& \vdots 		& \vdots	&\vdots		\\
e^{j(m-1)\psi_{1}} 	& e^{j(m-1)\psi_{2}} 	& \cdots 	& e^{j(m-1)\psi_{q}}
\end{array}\right)
\left( \begin{array}{cccc}
s_{1}(t_{1})	& s_{1}(t_{2})	& \cdots	& s_{1}(t_{M})	\\
s_{2}(t_{1})	& s_{2}(t_{2})	& \cdots	& s_{2}(t_{M})	\\
\vdots		& \vdots	& \vdots	& \vdots	\\
s_{q}(t_{1})	& s_{q}(t_{2})	& \cdots	& s_{q}(t_{M})	\\
\end{array}\right)&\\ \\
+\space \left(\begin{array}{cccc}
n_{z,1}(t_{1})	& n_{z,1}(t_{2})& \cdots	& n_{z,1}(t_{M})\\
n_{z,2}(t_{1})	& n_{z,2}(t_{2})& \cdots	& n_{z,2}(t_{M})\\
\vdots		& \vdots	& \vdots	& \vdots	\\
n_{z,m}(t_{1})	& n_{z,m}(t_{2})& \cdots	& n_{z,m}(t_{M})\\
\end{array}\right)&
\end{split}
\end{equation}
or
\begin{equation}
Z = A_{z}S + N_{z}
\end{equation}
where
\begin{equation*}
Z=\left(\begin{array}{cccc}
z_{1}(t_{1})	& z_{1}(t_{2})	& \cdots	& z_{1}(t_{M})	\\
z_{2}(t_{1})	& z_{2}(t_{2})	& \cdots	& z_{2}(t_{M})	\\
\vdots		& \vdots	& \vdots	& \vdots	\\ z_{m}(t_{1})	& z_{m}(t_{2})	& \cdots	& z_{m}(t_{M})	\\
\end{array}\right)
\end{equation*}
The matrices $A_{z}$ and $S$ are unknown, and are not rank-deficient by assumption.

Now, let $y_{1}=e^{j\psi_{1}}$, $y_{2}=e^{j\psi_{2}}$,..., $y_{q}=e^{j\psi_{q}}$ and assume that $y_{1}, y_{2},..., y_{q}$ are the roots of a polynomial equation
\begin{equation}
{\cal{P}}(y)=1+c_{1}y+c_{2}y^{2}+\cdots+c_{m-1}y^{m-1} = 0.
\label{polynomial}
\end{equation}
Now our aim is to calculate the coefficients which satisfy this assumption and then to solve for the roots.
\subsection{Method without Noise}
From (\ref{polynomial}) and (\ref{eq:Z_Data_Matrix}), it can easily be shown that the following equations hold  for no noise case.
\begin{eqnarray*}
z_{1}(t_1)+c_{1}z_{2}(t_1)+c_{2}z_{3}(t_1)+\cdots+c_{m-1}z_{m}(t_1)  &=&  0\\
z_{1}(t_2)+c_{1}z_{2}(t_2)+c_{2}z_{3}(t_2)+\cdots+c_{m-1}z_{m}(t_2)  &=& 0\\
\vdots\hspace{2in}& &\vdots  \\
z_{1}(t_M)+c_{1}z_{2}(t_M)+c_{2}z_{3}(t_M)+\cdots+c_{m-1}z_{m}(t_M) &=& 0 \end{eqnarray*}
After simple manipulations,
\begin{eqnarray*}
c_{1}z_{2}(t_1)+c_{2}z_{3}(t_1)+\cdots+c_{m-1}z_{m}(t_1) & =& -z_{1}(t_1) \\
c_{1}z_{2}(t_2)+c_{2}z_{3}(t_2)+\cdots+c_{m-1}z_{m}(t_2)  &=& -z_{1}(t_2)\\
\vdots\hspace{2in} & &\vdots  \\
c_{1}z_{2}(t_M)+c_{2}z_{3}(t_M)+\cdots+c_{m-1}z_{m}(t_M)  &=& -z_{1}(t_M) \end{eqnarray*}
The above set of equations can be written in matrix form as
\begin{equation}
\left(\begin {array}{cccc}
z_{2}(t_1) & z_{3}(t_1) & \cdots & z_{m}(t_1)\\
z_{2}(t_2) & z_{3}(t_2) & \cdots & z_{m}(t_2)\\
	 & \vdots   & 	     &\vdots \\
z_{2}(t_M) & z_{3}(t_M) & \cdots & z_{m}(t_M)
      \end {array}\right)
\left(\begin{array}{c}
c_{1}\\
c_{2}\\
\vdots\\
c_{m-1}
      \end{array}\right)=
\left(\begin{array}{c}
-z_{1}(t_1)\\
-z_{1}(t_2)\\
\vdots\\
-z_{1}(t_M)
      \end{array}\right)
\end{equation}
or
\begin{equation}
PC=P_{1}
\label{system}
\end{equation}
This is an overdetermined system of equations. The best solution in the least square sense for the coefficient vector C is given by
\begin{equation}
C=P^{\#}P_{1} 
\label{coefficients}
\end{equation}
where $P^{\#}$ is the Moore-Penrose pseudoinverse of $P$,
\begin{equation}
P^{\#}=(P^{H}P)^{-1}P^{H}\nonumber
\end{equation}
Substitute $c_{1},c_{2},\ldots,c_{m-1}$ obtained in (\ref{coefficients}) into (\ref{polynomial}) and solve for the roots of ${\cal{P}}(y)$. There are $(m-1)$ roots, and among those only q solutions are of our interest. From the $(m-1)$ roots choose $q$ roots say \{$\gamma_{l}$\}, whose magnitudes are nearer to unity:
\begin{equation}
\gamma_{l}=e^{j2\pi(d/\lambda)\cos{\theta_{l}}}\hspace{0.7in}l=1,2,\ldots,q
\end{equation}
Then, the estimate of $\psi_{l}$ is
\begin{equation}
\hat{\psi_{l}} = angle\{\gamma_{l}\}
\end{equation}
\\
Follow the similar approach for the X-subarray to calculate \{$\beta_{l}$\} defined by
\begin{equation}
\beta_{l}=e^{j2\pi(d/\lambda)\sin{\theta_{l}\cos{\phi_{l}}}}\hspace{.7in}l=1,2,\ldots,q 
\end{equation}
and hence
\begin{equation}
\hat{\xi_{l}}=angle\{\beta_{l}\}\hspace{1in}
\end{equation}
\\
Now the estimates of $\theta_{l}$ and $\phi_{l}$ are given by
\begin{equation}
\hat{\theta_{l}}=\sin^{-1}{\left(\frac{\hat\psi_{l}}{2\pi(d/\lambda)}\right)}\hspace{1in} l=1,2,\ldots,q.
\end{equation}
\begin{equation}
\hat{\phi_{l}}=\cos^{-1}{\left(\frac{\hat\xi_{l}}{(2\pi (d/\lambda) \sin{\hat\theta_{l}})} \right) }\hspace{.9in}l=1,2,\ldots,q.
\end{equation}

\subsection{Method with Noise}
Note that in (\ref{z_vector_in_sum}) and (\ref{x_vector_in_sum}), the vectors $\mathbf{n}_{z}(t)$ and $\mathbf{n}_{x}(t)$ are random noise components on Z and X subarrays respectively. The noise is assumed to be additive, temporally white Gaussian with zero mean. This assumption is a valid assumption in real-world applications.

Mathematically, we can express the above stated properties of the noise  as follows,
\begin{equation}
\label{eq:noise_propery_first}
E[\mathbf{n}_{z}(t)]=E[\mathbf{n}_{x}(t)]=\mathbf{0}
\end{equation}
\begin{equation}
E[\mathbf{n}_{z}(t)\mathbf{n}_{z}^{H}(t)]=E[\mathbf{n}_{x}(t)\mathbf{n}_{x}^{H}(t)] = \sigma^{2}\mathbf{I}
\end{equation}
\begin{equation}
E[\mathbf{n}_{z}(t)\mathbf{n}_{z}^{T}(t)]=E[\mathbf{n}_{x}(t)\mathbf{n}_{x}^{T}(t)] = \mathbf{0}
\end{equation}
and
\begin{equation}
E[\mathbf{n}_{z}(t_{i})\mathbf{n}_{z}^{H}(t_{j})]=E[\mathbf{n}_{z}(t_{i})\mathbf{n}_{z}^{T}(t_{j})] = \mathbf{0}\hspace{.5in}\text{for } i\neq j
\end{equation}
\begin{equation}
\label{eq:noise_property_last}
E[\mathbf{n}_{x}(t_{i})\mathbf{n}_{x}^{H}(t_{j})]=E[\mathbf{n}_{x}(t_{i})\mathbf{n}_{x}^{T}(t_{j})] = \mathbf{0}\hspace{.5in}\text{for } i\neq j
\end{equation}
where $E$ denotes the expectation operation, $^H$ denotes the Hermitian operation, $^T$ denotes the transposition, and \space$t_{i}$ and $t_{j}$ are the $i$th and $j$th time instants respectively.

The noise has an negative impact on the accuracy of estimates of the polynomial coefficients. In order to reduce the effect of the noise, we follow the same procedure except for the calculation of inverse of the matrix $P$ in (\ref{system}) where we use the singular value decomposition(SVD) as discussed below.\\
Let the SVD of $P$ be
\begin{equation}
P=U \Sigma V^{H}
\end{equation}
where $U$ and $V$ are the unitary matrices containing the singular vectors of P and $\Sigma$ is a diagonal matrix of same dimensions as $P$, and it is given by
\begin{equation}
\Sigma = U^{H}PV
\end{equation}
The diagonal elements in $\Sigma$ are non-negative and are sorted in decreasing order, i.e., 
\begin{equation}
\Sigma = diag\{\tau_{1},\tau_{2},\ldots,\tau_{q},\ldots,\tau_{m-1}\}
\end{equation}
where
\begin{equation}
\tau_{1}\geq\tau_{2}\geq\cdots\geq\tau_{q}\geq\cdots\geq\tau_{m-1}\nonumber
\end{equation}
are the singular values of $P$.

To reduce the effect of noise, replace the last $m-q-1$ singular values with zero in calculating the inverse of $P$, i.e.,
\begin{equation}
\Sigma_{1}=diag\{\tau_{1},\tau_{2},\ldots,\tau_{q},0,\ldots\}
\end{equation}
and
\begin{equation}
\Sigma_{1}^{-1}=diag\{1/\tau_{1},1/\tau_{2},\ldots,1/\tau_{q},0,\ldots\}
\end{equation}
The dimensions of $\Sigma_{1}^{-1}$ are same as that of $P^{H}$.\\
The lease squares approximation of $P^{-1}$ is given by ~\cite{Dewilde}
\begin{equation}
P^{-1}=V\Sigma_{1}^{-1}U^{H}
\end{equation}
hence, the estimate of coefficient vector in (\ref{system}) is
\begin{equation}
C=V\Sigma_{1}^{-1}U^{H}P_{1}
\end{equation}
Follow the same approach for X-subarray to obtain the coefficient  vector and the roots. The computation of AOAs from the roots is same as that for the noiseless case.

\end{document}